\documentclass[doublecol]{epl2} 
\usepackage{amssymb,amsmath}

\title{Correlation between Voronoi volumes in disc packings}

\shorttitle{Voronoi volume correlation} 

\author{Song-Chuan Zhao\inst{1}\thanks{E-mail: \email{songchuan.zhao@ds.mpg.de}} \and Stacy Sidle\inst{2} \and Harry L. Swinney\inst{2} \and Matthias Schr\"{o}ter\inst{1}\thanks{E-mail: \email{matthias.schroeter@ds.mpg.de}}}
\shortauthor{S. C. Zhao \etal}

\institute{                    
  \inst{1} Max-Planck-Institut f\"{u}r Dynamik und Selbstorganisation - Bunsenstr. 10, D-37073 G\"{o}ttingen, Deutschland\\
  \inst{2} Center for Nonlinear Dynamics and Department of Physics, University of Texas at Austin, Austin, Texas 78712, USA
}

\abstract{
We measure the two-point correlation of free Voronoi volumes in binary disc packings, where the packing fraction $\phi_{\rm avg}$ ranges from 0.8175 to 0.8380. We observe short-ranged correlations over the whole range of $\phi_{\rm avg}$ and  anti-correlations for $\phi_{\rm avg}>0.8277$.
The spatial extent of the anti-correlation increases with $\phi_{\rm avg}$ while the position of the maximum of the anti-correlation and the extent of the positive correlation shrink with $\phi_{\rm avg}$.  We conjecture that the onset of anti-correlation corresponds to dilatancy onset in this system. }

\begin{document}

\maketitle
\section{Introduction}
Dry granular matter interacts only via elastic and frictional forces, which require particles to be in contact;
spatially extended interactions like Van der Waals forces typically play no role. However,  
granular particles form networks of force chains \cite{majmudar:05}, which implies the existence of local correlations.
Lechenault {\it et al.} have shown that the logarithm of the distribution of free volumes\footnotemark\footnotetext{The free Voronoi volume is the difference between the actual Voronoi volume of a cell and the volume of a cell in a hexagonal packing} 
in granular matter scales in a non-extensive way with the cluster size, 
which implies the existence of correlations between Voronoi cells~\cite{lechenault:06} . 
A similar scaling was also observed in monodisperse sphere packings~\cite{aste:07_2}.   
These experimental observations raise the question how the spatial extent of such correlations 
changes with packing fraction. This question is especially important for granular systems with glassy behavior, 
where a number of groups have studied the length scale related to spatially heterogeneous 
dynamics~\cite{goldman:05,dauchot:05,aaron:07,lechenault:08}. 
In this paper we demonstrate the existence of both correlations and anti-correlation
in the free Voronoi volumes, and we measure their spatial extent as a function of volume fraction. 

\section{Experiment}
\begin{figure*}
\begin{center}
\includegraphics[width=1\textwidth]{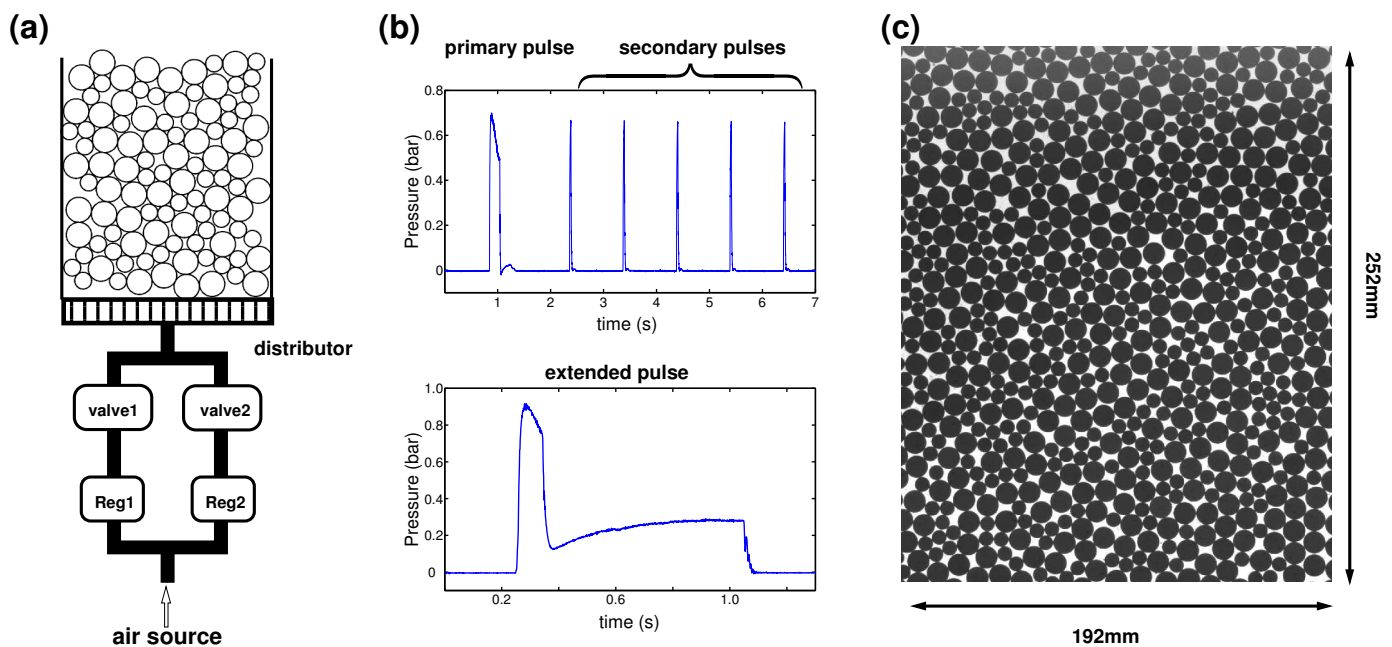}
\caption{ (a) Sketch of the experimental setup (particles not drawn to scale). (b) Examples of the different tapping modes. 
The pressure is  measured below  the distributor. (c) Image of the experimental packing. }
\label{f.setup}
\end{center}
\end{figure*}

Experiments are performed in a two-dimensional air fluidized bed, as sketched in fig.~\ref{f.setup}(a). The particles are a binary 
mixture of Teflon discs with diameters of $d_s = 6\un{mm}$ and $d_l = 9\un{mm}$. They are confined between two vertical glass plates 
(thickness $12\un{mm}$) separated by a distance slightly larger than the thickness of the discs ($3.86\un{mm}$). 
The bed contains approximately 750 discs of each size. 

The disc packing is tapped from below by air pulses  flowing through a distributor of open-porous foam (Duocel 40 PPI aluminium foam). 
Electrostatic charging is minimized by grounding the distributor.
The duration and strength of air pulses are controlled by two Waston Smith 06B04604 mechanical pressure regulators and two pairs 
of Jefferson 2026 series electronic valves.
Three sensors below the distributor are used to measure air pressure (Validyne DP15), humidity (Honeywell HIH-3610) and temperature (YSI 44033).
Typical humidities and temperatures are $3.8\pm0.3\%$ and $24.8\pm0.4^\circ C$.

The average packing fraction $\phi_{\rm avg}$ value is controlled by the type and duration of the air pulses. The use of different tapping modes (cf. fig.~\ref{f.setup}(b)) enables us to vary $\phi_{\rm avg}$  from 0.8175 to 0.8380,  as shown in table~\ref{t.par_air}. The most compact beds are obtained by first driving the bed to a new configuration using a primary pulse, and then following that by shorter secondary pulses to compactly the bed. The duration of the primary pulse and the strength of air flow are fixed, 
but the number and duration of the secondary pulses are adjustable. Looser packings are obtained using two airflow pathways with different flow rates; when the primary pulse stops, the secondary pathway still provides some small flow, which slows down the settling discs. 

\begin{table}
\caption{Parameters of the air pulses. All the experiments start with the same primary pulse ($3.2\un{bar}$   measured at the regulator, $200\un{ms}$), either followed by several secondary pulses with the same pressure or accompanied by an  extended pulse.}
\label{t.par_air}
\begin{center}
\begin{tabular}{c c c}
$\phi_{\rm avg}$	& Extended pulse	& Secondary pulses\\
0.8175	& $0.5\un{bar}\times1000\un{ms}$	& -\\
0.8183	& $0.45\un{bar}\times1000\un{ms}$	& -\\
0.8209	& -	& 0 \\
0.8218	& -	& $3\times15\un{ms}$ \\
0.8231	& -	& $5\times15\un{ms}$ \\
0.8256	& -	& $10\times15\un{ms}$ \\
0.8277	& -	& $3\times30\un{ms}$ \\
0.8315	& -	& $7\times30\un{ms}$ \\
0.8337	& -	& $15\times30\un{ms}$ \\
0.8356	& -	& $20\times30\un{ms}$ \\
0.8366	& -	& $15\times30\un{ms} + 15\times15\un{ms}$ \\
0.8380	& -	& $30\times30\un{ms} + 30\times15\un{ms}$ \\

\end{tabular}
\end{center}
\end{table}

After each tap, the packing is allowed to relax for four seconds. Then an image of the packing is taken by a CCD camera with a Nikkor 50mm lens. 
Only the central region of the packing ($252 \times 192\un{mm^2}$),  five small disc diameters away from boundaries, is captured. The spatial resolution is $0.17\un{mm}$ per pixel. 
For each experiment 8000 configurations are imaged.

\section{Image Processing}

To analyze the configuration we compute the Voronoi tessellation of the packing. 
Because we consider a 2D system, volume and area are used inter-changeably in the following.
In a first step the centers and sizes of the discs are calculated with an accuracy of 0.1 pixel 
using a template correlation technique.

\begin{figure*}
\begin{center}
\includegraphics[width=1\textwidth]{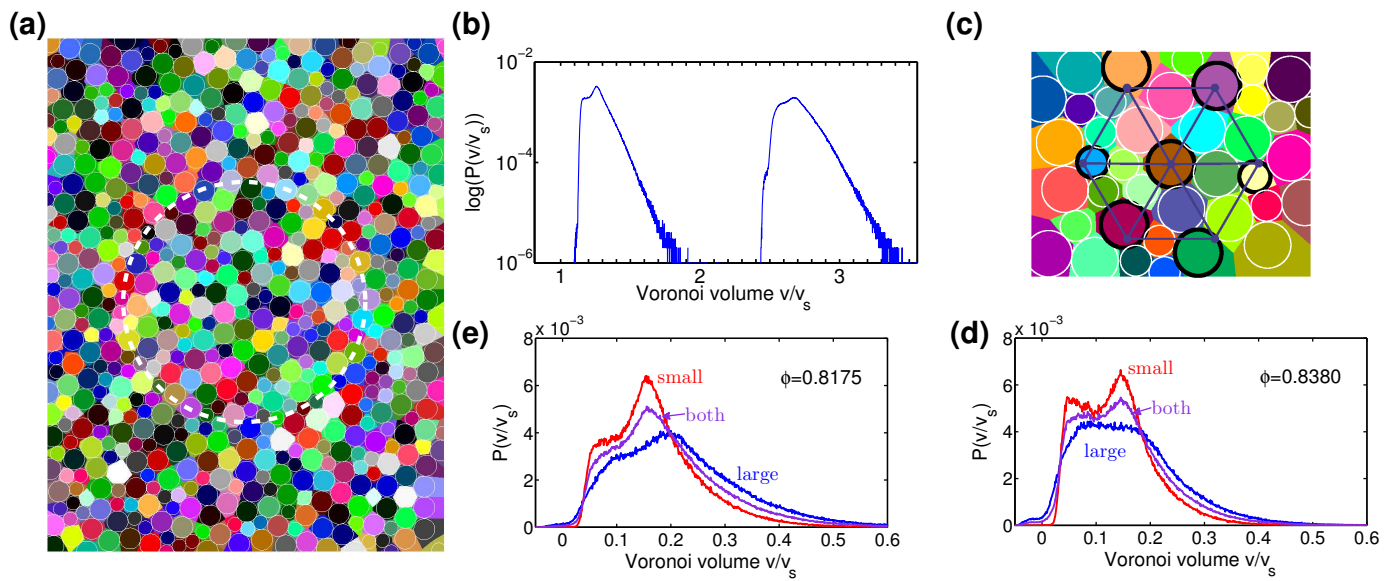}
\caption{ (a) Voronoi cells computed using the navigation map for the packing in  fig.~\ref{f.setup}(c) and labeled with different colors.  The white dashed circle indicates the size of the region over which we perform further analysis for over 8000 individual configurations. (b) The probability distribution of the volume of individual Voronoi cells; the two peaks correspond to the two sizes of discs. (c)  An illustration of the construction of pairs by a hexagonal wheel. The wheel vertices (dark dots) define 12 pairs of points. The free volume of the Voronoi cells to which the vertices belong is used for the two-point correlation measurement.  (d) and (e) Free volume distributions of small discs, large discs, and both for two different packing fractions.  The volume is normalized by the volume of small particles $v_s=\pi d_s^2/4$.  The probabilities are normalized by the number of small, large, and both discs, respectively.  }
\label{f.voronoi}
\end{center}
\end{figure*}

There are two methods to identify Voronoi cells in a bidisperse system, radical tessellation~\cite{radical:82} and navigation map \cite{richard:01}. 
Radical tessellation takes the boundaries of the cells as the collection of points whose tangents to neighboring particles are equal length. 
The navigation map takes the cells as the collection of points closer to the surface of the corresponding particles than others in the system. 
In this work the navigation map is computed numerically on a grid of 1/16 pixel resolution.
An example of the navigation map is shown in fig.~\ref{f.voronoi}(a).

Our packing is prepared under gravity.  Sidewalls introduce slow convection rolls during an air flow pulse, and distributor inhomogeneities introduce gradients in the airflow. It is therefore not surprising to find gradients of the packing fraction in the system\footnotemark\footnotetext{The spatial variance of the whole system is about  0.0435.}.
The analysis is done for the central spatially most homogeneous region. The spatial variance of the local packing fraction (averaged over the whole 8000 taps) is calculated for circles of diameter 18$d_s$ {(see the white dashed circle in figure \ref{f.voronoi}(a))}.   
Then the region with the smallest variance (in all cases smaller than {0.0043}) is chosen as the analysis region for the experiment.
Additionally, the evolution of the global packing fraction and the geometrical contact number in that region are examined 
to make sure that no segregation occurs during the course of the experiment.

\section{Results}
\subsection{Individual Voronoi volumes and free volume distribution}
The distribution of individual Voronoi volumes $v$ can be computed directly from the results of the navigation map. An example
 is shown in fig.~\ref{f.voronoi}(b). 
Because we do not want to distinguish between small and large discs, we follow Lechenault et al.~\cite{lechenault:06}
and use in the subsequent analysis the free Voronoi volume $v_f = v - v_{min}$. 
 The minimum volume $v_{min}$ is the volume that a grain would occupy in a hexagonal packing of identical discs. 
It equals $\frac{\sqrt{3}}{2}d^2$, where 
$d$ is the diameter of the corresponding large ($d_l$) discs or small ($d_s$) discs. While the respective mean free volume 
of small and large discs still differ ($\bar{v}^f_s=0.135$ and $\bar{v}^f_l=0.158$  in units of $v_s = \pi d_s^2/4$ for $\phi_{\rm avg}=0.8380$), 
the success of the subsequent analysis justifies this step by hindsight. 

The free volume distributions of small and large discs and both types together are presented for two packing fractions
 in fig.~\ref{f.voronoi}(d) and (e). For large volumes the decay is exponential but besides this feature none of the free volume distributions and volume distributions 
in fig.~\ref{f.voronoi}(b),  (d) and (e) could be fitted reasonably with a gamma distribution. 
This result differs from 2D and 3D mono-disperse packings \cite{aste:07,b.gamma_dist2,kumaran:05}.

The relative height of the two peaks in the free volume distributions changes  with $\phi_{\rm avg}$. 
A recent study of the probability distributions of quadron volumes ({an alternative tessellation} introduced in \cite{raphael:03}) 
showed that the position and height of these peaks can be traced to conditional probabilities of cell volumes at given 
coordination numbers~\cite{b.shoulder}. 
So the changes visible in figure \ref{f.voronoi}(d) and (e) may also be related to a change of contact number. 

\subsection{Two-point correlation of free Voronoi volume}
To quantify the correlations between Voronoi volumes we use the two-point correlation function:

\begin{equation}
\label{eq.t_corr}
C_{ij}(L) = \frac{\langle(v_{i}-\bar{v}_{i})(v_{j}-\bar{v}_{j})\rangle}{\sigma^2}
\end{equation}
where $i$, $j$ correspond to two points  in a distance $L$ belonging to two Voronoi volumes, and  $v_i$, $v_j$ are the free volume of these Voronoi cells. 
$\langle...\rangle$ indicates averages over all the 8000 different packings created by  flow pulses. 
$\bar{v}_i$ and $\bar{v}_j$ are the mean free volumes at these points
 (computed separately to remedy the effect of remaining small gradients), and
$\sigma^2 = (\sigma_i^2+\sigma_j^2) /2$  is the corresponding variance.

To obtain better statistics, 240 pairs of points with the same $L$ are selected. For each pair $C_{ij}(L)$ is computed
using eq.~\ref{eq.t_corr}. In practice the pairs are selected in 
the following way: 
We construct a hexagonal ``wheel'' centered in the analysis region. The length of each edge of the wheel 
is set to be $L$ (c.f.~fig.~\ref{f.voronoi}(e)). The six outer points and the center form 12 pairs that are separated by a distance $L$. 
For each such pair, the free volume of the two Voronoi cells to which the two vertices belong are taken as $v_i$ and $v_j$ in eq.~\ref{eq.t_corr}. 
Then the hexagon is translated to five different positions and rotated to four different angles for each position. {However, the whole wheel stays inside the white circle in fig.~\ref{f.voronoi}(a).}
Then the 240 pairs are averaged:

\begin{equation}
\label{eq.corr}
corr(L)=\frac{1}{240}\sum_{i,j}C_{ij}(L).
\end{equation}

\begin{figure}
\onefigure[width=0.49\textwidth]{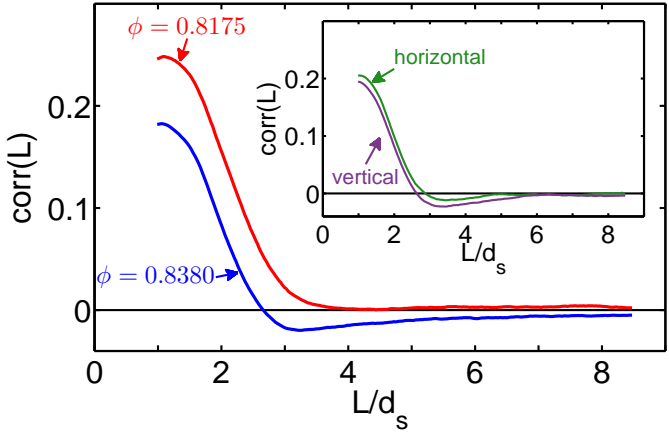}
\caption{Depending on $\phi$,  $corr(L)$ exhibits both correlations and anti-correlations. Inset: the anisotropy measurement for $\phi_{\rm avg}=0.8380$. The two point correlation of pairs placed horizontally (dark green) and vertically (brown), rather than constructed from the hexagon.}
\label{f.corr_sample}
\end{figure}

The correlation function $corr(L)$ is shown for two values of $\phi_{\rm avg}$ in fig.~\ref{f.corr_sample}. For low $\phi_{\rm avg}$, $corr(L)$ 
decays to zero and then fluctuates around it.
Positive values of $corr(L)$ indicate that the two free Voronoi volumes deviate from the average in the same direction.
For high $\phi_{\rm avg}$, $corr(L)$ decreases to a negative minimum and then increases towards 0. Negative values of $corr(L)$ characterize 
anti-correlations: Voronoi cells at this distance deviate in opposite directions from the average free volume. 

\begin{figure}
\onefigure[width=0.49\textwidth]{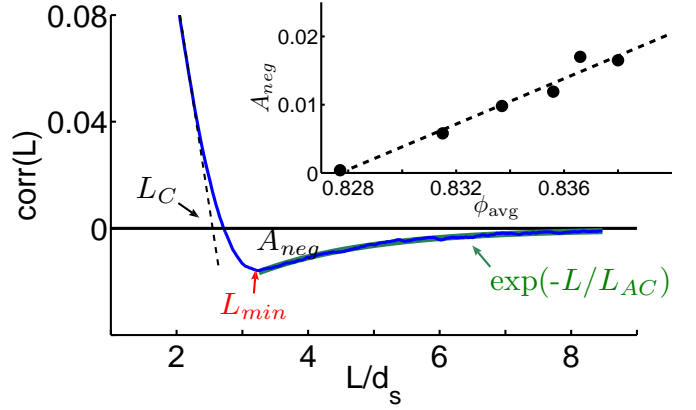}
\caption{Three characteristic lengths: $L_C$, obtained from a linear fit (dashed black line) to $corr(L)$ for small $L$;  $L_{min}$, corresponding to the minimum of $corr(L)$; and $L_{AC}$, obtained from an exponential fit (green line) to $corr(L)$ for $L > L_{min}$.  The inset shows the negative area $A_{neg}$ of $corr(L)$ as a function of of $\phi_{avg}$.}
\label{f.threshold}
\end{figure}

We define three characteristic lengths to describe $corr(L)$ (see fig.~\ref{f.threshold}). 
First we do a linear fit to  $corr(L)$ for a range centered at half of the 
maximum of $corr(L)$ with width $\pm 1/2 d_s$. The point where this fit crosses zero yields the length $L_C$. 
For measurements of $corr(L)$ showing anti-correlation we define
a second length, $L_{min}$, which is extracted from a local parabolic fit around the minimum of $corr(L)$. 
The third length, $L_{AC}$, is obtained from an exponential fit ranging from $L_{min}$ to the end of $corr(L)$ (we allow for a small offset in this fit, but the magnitude of this offset is less than 0.0006 in all cases). 

To find the  onset of the anti-correlations, the area $A_{neg}$  where $corr(L)<0$ (cf. fig.~\ref{f.threshold}) is plotted as a function of $\phi_{avg}$, as shown in inset of fig.~\ref{f.threshold}~\cite{offset}.
An extrapolation of a linear fit to $A_{neg}$ to zero defines the threshold for anti-correlations, $\phi_{AC}=0.8277\pm0.0005$. 
For $\phi>\phi_{AC}$, $corr(L)$ can be both positive and negative; therefore, it can not be described by a power law~\cite{lechenault:06}.

\begin{figure}
\onefigure[width=0.49\textwidth]{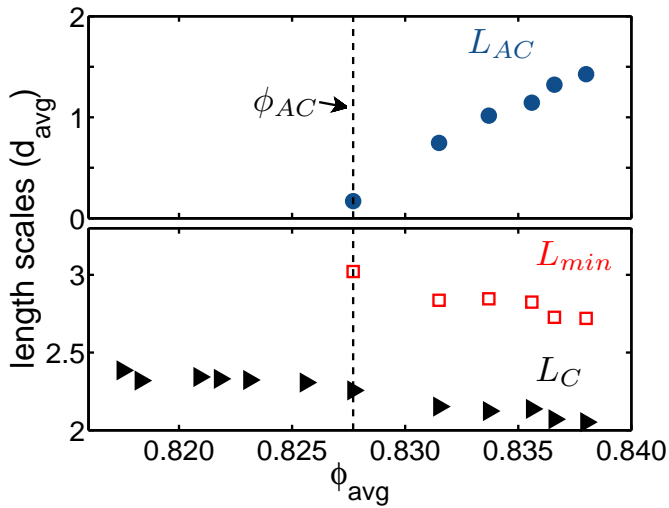}
\caption{Packing fraction dependence of the characteristic lengths  of correlation $L_C$ and anti-correlation $L_{min}$ and $L_{AC}$.  The lengths are normalized using the average disc diameter $d_{avg}$, which corresponds to $1.23-1.28$ times $d_s$.}
\label{f.lengthscale}
\end{figure}

\begin{figure}
\onefigure[width=0.4\textwidth]{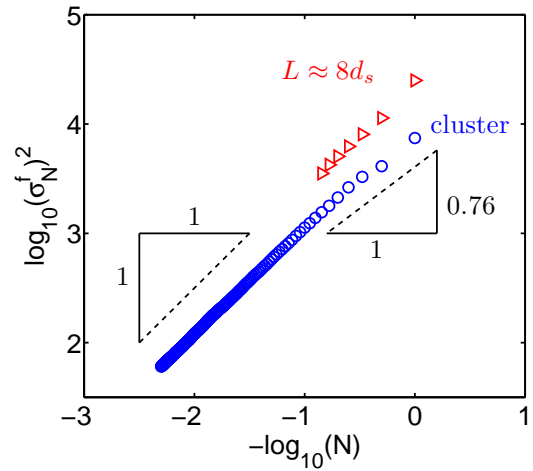}
\caption{The correlations have finite spatial extent, as shown by
the variance ${\sigma_N^f}^2$ of the average free volume of a compact cluster as a function of the number $N$ of grains included  
 (open blue circles). In the absence of correlations the central limit theorem predicts a slope of one. While the slope for small clusters ($N < 10$) is 0.76, in the large $N$ (and therefore large $L$) limit a slope of one is recovered. Also the slope of 
of a hexagonal wheel ``cluster'' with side length $L \approx 8d_s$ (open triangles) equals one. 
 Here $\phi_{\rm avg}=0.8380$; data sets are shifted for clarity.}
\label{f.scaling}
\end{figure}

The dependence of  $L_C$, $L_{min}$, and $L_{AC}$ on $\phi_{\rm avg}$ is shown in figure \ref{f.lengthscale}:  $L_C$ and $L_{min}$ slowly decrease 
monotonically with $\phi_{\rm avg}$,  while $L_{AC}$ grows approximately linearly. The finite extent of correlations and anti-correlations can also be seen in in fig.~\ref{f.scaling}, which shows how the variance of the free volume of a cluster scales with the number $N$ of particles in the cluster.
In the large $N$ limit a linear relationship is recovered as predicted by the central limit theorem in the absence of
correlations.   

Gravity breaks the isotropy in our experimental setup; this anisotropy is visible in the
correlations plotted in the inset of figure \ref{f.corr_sample}, where pairs
of horizontal and vertical points are averaged separately. While the qualitative features of
$corr(L)$ are independent of direction, the lower statistics of this analysis 
does not allow us to extract the corresponding characteristic lengths.

We have also performed our analysis with the radical tessellation and have found that all
features stay qualitatively the same. Quantitatively there are slight differences: 
$\phi_{AC}=0.828\pm0.002$; $L_C$, $L_{min}$ and $L_{AC}$ are $6\%, 12\%$ and $30\%$ larger on average.

\section{Discussion}
This is the first observation of anti-correlation between Voronoi volumes in a granular system.  One reason for this is that earlier 
studies of volume fluctuations focused on the cluster composed of neighboring particles~\cite{lechenault:06,aste:07_2}. There the 
variance of average free volume per grain $\sigma_N^2$ as a function of the number of grains included $N$ was measured, and the scaling $\sigma_N^2\sim N^{-\alpha}$ with $\alpha<1$ indicated the existence of correlations.  However, the scaling between $\sigma_N^2$ and $N$ is a measure of the correlation integrated over the whole cluster size. 
In the large $N$ limit, the relation between $N^{-\alpha}$ and $corr(L)$ is:
\begin{equation}
\label{eq.alpha_scaling}
 N^{-\alpha}=\frac{\mathcal{C}}{N}\int_0^{L(N)} corr(L)L\upd L
\end{equation}
where $\mathcal{C}$ is a constant proportional to the density of the cluster ($\pi L(N)^2$ would be the size of the cluster). 
The relatively small contribution of anti-correlations to the integral in eq.~\ref{eq.alpha_scaling} would be hard to distinguish 
from experimental noise (cf. fig.~\ref{f.scaling}). Therefore, a direct measurement of two-point correlation is necessary to find anti-correlation.

In recent years a statistical mechanics approach for granular systems has been developed where the Hamilton function is replaced by a volume function \cite{edwards:89}. 
Voronoi cells have been used to construct this function~\cite{lechenault:06,aste:08,makse:nature08}. 
While the discovery of anti-correlation has consequences for such an approach, some properties like the distribution of the local packing fraction have been shown to not depend on correlations~\cite{puckett:11}.

It is possible that not only the extent but even the existence of positive correlation is an artifact of  Voronoi tessellation. {By definition} the Voronoi tessellation assigns space to each particle in a certain `equal' way. In the case of the navigation map, the edge of a Voronoi cell is the collection of points that have equal distance $r$ to the surface of the neighboring particles. The increase in a Voronoi volume could roughly be seen as the increase in $r$, therefore the $r$ of its neighboring Voronoi cell. Consequently the Voronoi tessellation itself gives rise to a positive correlation between neighboring Voronoi volumes. To verify whether this effect dominates the evolution of $L_C$, we estimate the average size $d_{\rm vor}$ of Voronoi cells in the system in the following way. In a homogeneous system, $\phi_{\rm avg}$ could be written as the ratio of the average particle volume to the average Voronoi volume, $\phi_{\rm avg} = d_{\rm avg}^2/d_{\rm vor}^2$. Given the bidisperse nature of our system $d_{\rm avg}$ is in the range of $1.23-1.28 d_s$. This estimation leaves us $d_{\rm vor} =   d_{avg}/\sqrt{\phi_{\rm avg}}$. Rescaling $L_C$ with $d_{\rm vor}$ shows that $L_C$ decays slightly faster than explained by simple compaction. This could stem from the appearance of the anti-correlation beyond $\phi_{AC}$. It will be interesting to test this hypothesis with another way of tessellation, such as quadrons~\cite{raphael:03}.

Above $\phi_{AC}$ the fluctuation of the volume of one grain causes more and more grains to be anti-correlated, which is indicated by the  increase of $L_{AC}$ and the decrease of $L_{min}$. Concerning the physical interpretation 
of {these} anti-correlations, we conjecture that $\phi_{AC}$ might correspond to the onset of dilatancy. In 3D systems it is well established that loose granular material collapses under shear, while granular material denser than dilatancy onset expands~\cite{kabla:prl09,gravish:prl10}. This expansion can be understood as competition between the grains for free volume.

{Our results show that correlations between the volumes of subunits depend on the specific system under consideration. For example, in froths there exist anti-correlations of the number of faces between neighboring cells~\cite{donovan:11,oger:96,hilhorst:08,kumar:94}. At the same time the volume of a foam cell is proportional to the number of faces~\cite{oger:96,stoyan:95}.  These features combined indicate that the volume of {\it neighboring} cells should be anti-correlated.}

It would be interesting to know how $L_{AC}$ would develop at even higher density.  
However, we can not compactify the system above $\phi=0.838$ without partial segregation. 
In this case no anti-correlations occur within the segregated patches. 
The only surviving length is $L_C$, which characterizes the size of the segregated patches.

\section{Conclusion}  
Our measurements of the two-point correlation function for binary disc packings yield a short-ranged positive correlation over the whole range of packing fractions $0.8175-0.8380$.
Above a volume fraction of $0.8277\pm0.0005$ we observe anti-correlation in the free Voronoi volumes. These anti-correlations reach a maximum at a distance of about 3.5 small particle diameters. They then decay exponentially with distance, with an exponent growing linearly with packing fraction.

\acknowledgments
We thank Sabrina Nagel, Udo Krafft and Wolf Keiderling for their technical support.

\bibliographystyle{eplbib}
\bibliography{2D_packing}

\begin{thebibliography}{10}
\bibitem{offset}
When computing $A_{neg}$ we take in to account the offset of the exponential
  fit determining $L_{AC}$.
\end{thebibliography}

\end{document}